\documentclass[twocolumn, twocolappendix, times, tighten]{aastex631}

\usepackage{savesym}
\savesymbol{tablenum}
\usepackage{siunitx}
\restoresymbol{SIX}{tablenum}

\begin{document}

\title{Oxygen Isotope Exchange Between Dust Aggregates and Ambient Nebular Gas}

\correspondingauthor{Sota Arakawa}
\email{arakawas@jamstec.go.jp}

\author[0000-0003-0947-9962]{Sota Arakawa}
\affiliation{Center for Mathematical Science and Advanced Technology, Japan Agency for Marine-Earth Science and Technology, 3173-25 Showa-machi, Kanazawa-ku, Yokohama 236-0001, Japan}

\author[0000-0001-6852-2954]{Daiki Yamamoto}
\affiliation{Department of Earth and Planetary Sciences, Kyushu University, 744 Motooka, Nishi-ku, Fukuoka 819-0395, Japan}

\author[0009-0005-8583-9730]{Lily Ishizaki}
\affiliation{Department of Earth and Planetary Science, The University of Tokyo, 7-3-1 Hongo, Bunkyo-ku, Tokyo 113-0033, Japan}

\author[0000-0003-1545-2723]{Tamami Okamoto}
\affiliation{Earth-Life Science Institute, Tokyo Institute of Technology, 2-12-1 Ookayama, Meguro-ku, Tokyo 152-8550, Japan}

\author[0000-0001-6502-6488]{Noriyuki Kawasaki}
\affiliation{Department of Earth and Planetary Sciences, Faculty of Science, Hokkaido University, Kita-10 Nishi-8, Kita-ku, Sapporo 060-0810, Japan}



\begin{abstract}

Meteorites and their components exhibit a diverse range of oxygen isotope compositions, and the isotopic exchange timescale between dust grains and ambient gas is a key parameter for understanding the spatiotemporal evolution of the solar nebula.
As dust grains existed as macroscopic aggregates in the solar nebula, it is necessary to consider the isotopic exchange timescales for these aggregates. 
Here, we theoretically estimate the isotope exchange timescales between dust aggregates and ambient vapor. 
The isotope exchange process between aggregates and ambient vapor is divided into four processes: (i) supply of gas molecules to the aggregate surface, (ii) diffusion of molecules within the aggregate, (iii) isotope exchange on the surface of constituent particles, and (iv) isotope diffusion within the particles. 
We evaluate these timescales and assess which one becomes the rate-determining step. 
We reveal that the isotope exchange timescale is approximately the same as that of the constituent particles when the aggregate radius is smaller than the critical value, which is a few centimeters when considering the exchange reaction between amorphous forsterite aggregates and water vapor.

\end{abstract}



\section{Introduction}
\label{sec:intro}

Oxygen is the most fundamental element for solids in the universe.
\citet{1973Sci...182..485C} discovered a large oxygen isotopic anomaly in refractory inclusions of the Allende carbonaceous chondrite, and oxygen isotopic compositions of extraterrestrial materials have been intensively investigated in the fields of meteoritics and planetary science \citep[e.g.,][]{2008Sci...321.1664N, 2022SciA....8E2067K, 2024arXiv240412536L}.

Meteorites and their components exhibit a diverse range of oxygen isotope compositions \citep[e.g.,][]{2013GeCoA.101..302S, 2017GeCoA.201...83K, 2018E&PSL.496..132M}.
This isotopic variation is usually interpreted as the result of mixing between isotopically distinct reservoirs in the solar nebula \citep[e.g.,][]{2012GeCoA..90..242U, 2018GeCoA.221..318K, 2018crpd.book..192T, 2022GeCoA.336..104Y}.
The reservoirs would be formed through self-shielding of carbon monoxide (CO) gas in the protosolar molecular cloud \citep[e.g.,][]{2004Sci...305.1763Y, 2020SciA....6.2724K} or the early solar nebula \citep[e.g.,][]{2005Natur.435..317L}.
Minor CO isotopologues (i.e., C$^{17}$O and C$^{18}$O) are selectively dissociated by irradiation of ultraviolet photons and $^{16}$O-poor oxygen atoms are produced.
They are subsequently transformed into $^{16}$O-poor water (H$_{2}$O) ice through chemical reactions at the surface of dust grains \citep[e.g.,][]{2004Sci...305.1763Y}.
Considering the mass balance, the residual CO gas would be enriched in $^{16}$O.
As a result, the three major oxygen reservoirs (i.e., H$_{2}$O, CO, and silicate) would have distinct isotope compositions \citep[e.g.,][]{2022ApJ...926..148F, 2023ASPC..534.1075N}.
The material evidence for the presence of $^{16}$O-poor H$_{2}$O reservoir in the solar nebula is preserved as cometary H$_{2}$O ice \citep[e.g.,][]{2019ARA&A..57..113A} and $^{16}$O-poor magnetite, formed by oxidation by H$_{2}$O, in ``cosmic symplectite'' found in the most primitive carbonaceous chondrite Acfer 094 \citep[e.g.,][]{2007Sci...317..231S}.

In a protoplanetary disk, dust grains drift radially due to gas drag and turbulent diffusion, and water ice sublimation occurs around the H$_{2}$O snowline \citep[e.g.,][]{2022ApJ...928..171O}.
Inward migration of icy dust grains could cause local enhancement of $^{16}$O-depleted H$_{2}$O relative to $^{16}$O-enriched CO \citep[e.g.,][]{2004ApJ...614..490C, 2005ApJ...622.1333K}.
Both H$_{2}$O and CO vapors could react with silicate grains in the inner region of the disk where the temperature is high \citep[e.g.,][]{2018ApJ...865...98Y, 2020M&PS...55.1281Y, YAMAMOTO202493, 2023ApJ...957...47I}, leading to depletion/enrichment of silicate grains in $^{16}$O.
Evaporation of silicate grains followed by recondensation also alters their oxygen isotopic composition \cite[e.g.,][]{2004GeCoA..68.3943A, 2012M&PS...47.1209N}.

\citet{2011Sci...332.1528M} reported the oxygen isotopic composition of the Sun, which represents the average isotopic composition of the solar system.
They found that the vast majority of silicate grains in the inner solar system were depleted in $^{16}$O compared to the composition of the Sun.
Assuming that silicate grains initially have a Sun-like isotope composition on average, the oxygen isotopic composition of silicate grains must evolve before the accretion of planetesimals.
The isotopic exchange between pristine silicate dust components with the Sun-like oxygen isotopic compositions and $^{16}$O-depleted H$_{2}$O vapor is a key process for the oxygen isotopic evolution in the solar nebula.

Dust grains in the interstellar medium are largely amorphous \citep[e.g.,][]{2004ApJ...609..826K}.
In contrast, astronomical observations at infrared wavelengths have revealed that both crystalline and amorphous silicate grains exist in protoplanetary disks, with forsterite being one of the most abundant crystalline silicate \citep[e.g.,][]{2009A&A...507..327O, 2010ApJ...721..431J}.
Therefore, we expect that amorphous forsterite represents the dust that existed in the early phase of the solar nebula.

The isotopic exchange timescale between dust grains and ambient gas is the key parameter to understanding the spatiotemporal evolution of oxygen isotopic composition of the solar nebula.
Therefore, laboratory experiments of oxygen isotope exchange reaction under the disk-like low vapor pressure conditions are essential.
In this context, \citet{2018ApJ...865...98Y} performed experiments of oxygen isotope exchange reaction between amorphous forsterite grains and water vapor and determined the isotopic exchange timescale.
The timescales for different minerals and gases have also been investigated \citep[e.g.,][]{2020M&PS...55.1281Y, YAMAMOTO202493}.
They found that, for amorphous forsterite grains with a radius of 1 micron or smaller, the timescale for oxygen isotope exchange at temperatures above approximately 600 K would be shorter than the lifetime of the solar nebula.
Additionally, at temperatures below 800 K, isotope exchange occurs faster than forsterite crystallization.

However, dust grains exist as aggregates in protoplanetary disks \citep[e.g.,][]{2023ApJ...944L..43T} and the solar nebula \citep[e.g.,][]{2016Natur.537...73B}, necessitating consideration of aggregates' isotopic exchange timescales in discussions on the evolution of oxygen isotope compositions in the solar nebula.
To quantify the effect of aggregate structure on the isotopic exchange timescale, understanding of the diffusion of gas molecules within the aggregate through its voids is essential.
Recently, the diffusion of dilute gas within aggregates have been intensively investigated in the context of cometary science \citep[e.g.,][]{2023MNRAS.524.6114G, 2024MNRAS.52712268S}.
We can employ theoretical models developed in the cometary science community for isotope exchange reaction of dust aggregates in the gaseous solar nebula.
In addition, the surface-to-volume ratio decreases as dust grains grow into larger aggregates, potentially affecting the isotopic exchange timescale by controlling the supply of molecules at the surface of aggregates.
We also quantify this effect and derive the critical aggregate radius below which the aggregate effects can be neglected.

In this study, we theoretically estimate the oxygen isotope exchange timescales between silicate dust aggregates and ambient water vapor.
The isotope exchange process between aggregates and ambient vapor is divided into four processes (see Section \ref{sec:model}).
We evaluate the timescales of these processes and assess which one becomes the rate-determining step based on aggregate size, temperature, and water vapor pressure.
Our analytical calculations reveal that, for dust aggregates smaller than cm size, the isotope exchange timescale is approximately the same as that of the constituent particles (see Section \ref{sec:results}).

\section{Model}
\label{sec:model}

\begin{figure}
\includegraphics[width = 0.6\columnwidth]{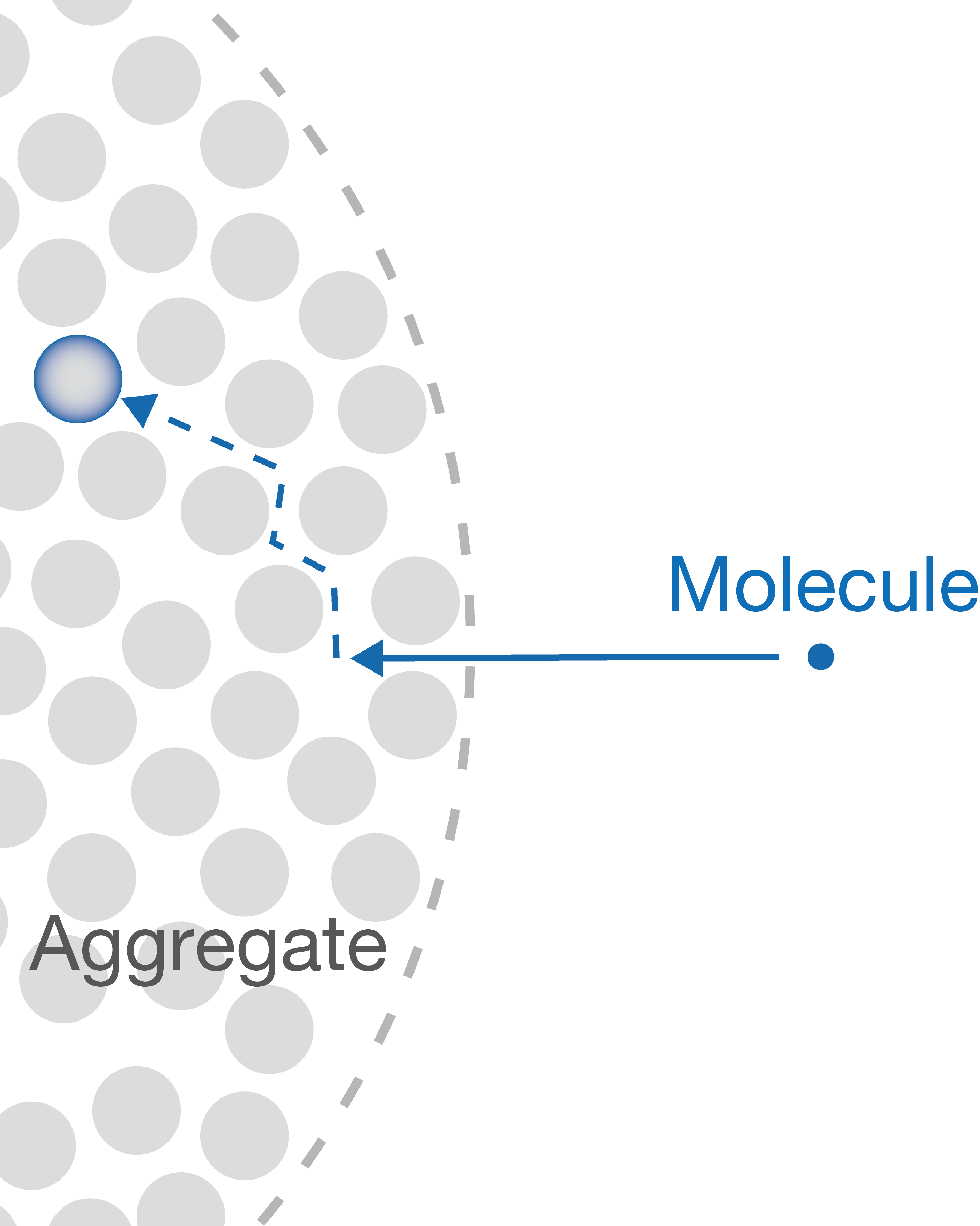}
\caption{
Schematic of the diffusion of gas molecules within a dust aggregate.
First, molecules are supplied to the surface of the aggregate from the ambient gas.
Some of these molecules are trapped within the aggregate, while others reflect off the aggregate surface without undergoing an exchange reaction.
The trapped molecules then diffuse through the voids within the aggregate and react with the surface of the constituent particles.
Finally, the isotopic composition of the constituent particles evolves through the diffusive exchange reaction within the particles.
}
\label{fig_schematic}
\end{figure}

Here, we construct a theoretical model of isotope exchange process between aggregates and ambient vapor.
Figure \ref{fig_schematic} shows a schematic of gas molecule diffusion within a dust aggregate.
The isotope exchange process between aggregates and ambient vapor is divided into four processes: (i) supply of gas molecules to the aggregate surface, (ii) diffusion of molecules within the aggregate, (iii) isotope exchange on the surface of constituent particles, and (iv) isotope diffusion within the particles.
The timescales for these four processes are denoted as $t_{\rm agg, surf}$, $t_{\rm agg, diff}$, $t_{\rm par, surf}$, and $t_{\rm par, diff}$, respectively.
We theoretically derive the equations for these four timescales and their parameter dependences in the following sections.

We evaluate the timescale for isotope exchange reaction, $t_{\rm ex}$, as the sum of the timescales of the four processes:
\begin{equation}
t_{\rm ex} = t_{\rm agg, surf} + t_{\rm agg, diff} + t_{\rm par, surf} + t_{\rm par, diff}.
\end{equation}
The isotope exchange timescale for constituent particles is given by
\begin{eqnarray}
t_{\rm ex, par} & =       & t_{\rm par, surf} + t_{\rm par, diff} \nonumber \\
                & \approx & \max{( t_{\rm par, surf}, t_{\rm par, diff} )}.
\end{eqnarray}
Thus, the effect of aggregate structure on the isotopic exchange timescale is significant when the following condition is satisfied:
\begin{equation}
\max{( t_{\rm agg, surf}, t_{\rm agg, diff} )} \gtrsim t_{\rm ex, par}.
\end{equation}

\subsection{Supply of Molecules at the Surface of Aggregates}

For a spherical aggregate with a radius of $r_{\rm agg}$, the timescale of the supply-controlled exchange reaction, $t_{\rm agg, surf}$, is given by \citep[e.g.,][]{2018ApJ...865...98Y}
\begin{equation}
t_{\rm agg, surf} = \frac{N_{\rm agg}}{\beta_{\rm agg} J S_{\rm agg}},
\label{eq:t_agg_surf}
\end{equation}
where $S_{\rm agg} = 4 \pi {r_{\rm agg}}^{2}$ is the surface area of the aggregate, $N_{\rm agg}$ is the number of oxygen atoms in the aggregate, $\beta_{\rm agg}$ is the exchange efficiency of molecules at the surface of the aggregate, and $J$ is the supply flux of oxygen atoms.
Here, $J$ is given by
\begin{equation}
J = \frac{P \gamma_{\rm gas}}{\sqrt{2 \pi m k_{\rm B} T}},
\end{equation}
where $P$ is the partial pressure, $m$ is the molecular weight, $T$ is the temperature, and $k_{\rm B} = 1.38 \times 10^{-23}~\si{J.K^{-1}}$ is the Boltzmann constant.
For water vapor, we set $m = 18 m_{\rm H}$, where $m_{\rm H} = 1.67 \times 10^{-27}~\si{kg}$ is the mass of a hydrogen atom.
We define the number of oxygen atoms per gas molecule as $\gamma_{\rm gas}$, and $\gamma_{\rm gas} = 1$ for H$_{2}$O.

The number of oxygen atoms in the aggregate, $N_{\rm agg}$, is given by
\begin{equation}
N_{\rm agg} = \frac{\phi_{\rm agg} V_{\rm agg} N_{\rm A} \gamma_{\rm grain}}{\Omega},
\end{equation}
where $\Omega$ is the molar volume of minerals, $V_{\rm agg} = {( 4 \pi / 3 )} {r_{\rm agg}}^{3}$ is the volume of the aggregate, $\phi_{\rm agg}$
is the volume filling factor of the aggregate, and $N_{\rm A} = 6.02 \times 10^{23}~\si{mol^{-1}}$ is the Avogadro constant.
We set $\Omega = 50~\si{cm^{3}.mol^{-1}}$ for both amorphous and crystalline forsterite \citep{2018ApJ...865...98Y}.
We define the number of oxygen atoms in the compositional formula of grains as $\gamma_{\rm grain}$, and $\gamma_{\rm grain} = 4$ for amorphous/crystalline forsterite (Mg$_{2}$SiO$_{4}$) grains.

The area fraction of holes on the surface of aggregates is equal to the porosity of aggregates, $1 - \phi_{\rm agg}$.
As exchange of molecules between aggregates and ambient gas occurs at the site of the hole on the surface of aggregates (see Figure \ref{fig_schematic}), the exchange efficiency of molecules at the surface of the aggregate, $\beta_{\rm agg}$, is approximately given by
\begin{equation}
\beta_{\rm agg} = 1 - \phi_{\rm agg}.
\end{equation}

\subsection{Diffusion of Molecules Within Aggregates}

For a spherical aggregate with a radius of $r_{\rm agg}$, the timescale of the diffusive exchange reaction, $t_{\rm agg, diff}$, is given by \citep[e.g.,][]{crank1979mathematics}
\begin{equation}
t_{\rm agg, diff} = \frac{{r_{\rm agg}}^{2}}{\pi^{2} D_{\rm agg}},
\label{eq:t_agg_diff}
\end{equation}
where $D_{\rm agg}$ is the diffusion coefficient within an aggregate.
For dust aggregates consisting of micron-sized grains, the size of voids is orders of magnitude smaller than the mean free path of gas molecules, and the random motion of gas molecules is governed by collisions with the particle surface \citep{1909AnP...333...75K}.
In this situation, $D_{\rm agg}$ is given by \citep[e.g.,][]{derjaguin1946flow, 2023MNRAS.524.6114G, 2024MNRAS.52712268S}
\begin{equation}
D_{\rm agg} = \frac{4 r_{\rm par} c_{\rm s} {( 1 - \phi_{\rm agg} )}^{2}}{13 \phi_{\rm agg}},
\end{equation}
where $r_{\rm par}$ is the radius of constituent particles and $c_{\rm s} = \sqrt{{( 8 k_{\rm B} T )} / {( \pi m )}}$ is the mean thermal velocity of molecules.

In Section \ref{sec:results}, we show that $t_{\rm agg, diff}$ for cm-sized aggregates is orders of magnitude shorter than the others ($t_{\rm agg, surf}$, $t_{\rm par, surf}$, and $t_{\rm par, diff}$).
Thus, the diffusion within aggregates is not the rate-limiting process.

\subsection{Supply of Molecules at the Surface of Constituent Particles}

For a spherical particle with a radius of $r_{\rm par}$, the timescale of the supply-controlled exchange reaction, $t_{\rm par, surf}$, is given by \citep[e.g.,][]{2018ApJ...865...98Y}
\begin{equation}
t_{\rm par, surf} = \frac{N_{\rm par}}{\beta_{\rm par} J S_{\rm par}},
\label{eq:t_par_surf}
\end{equation}
where $S_{\rm par} = 4 \pi {r_{\rm par}}^{2}$ is the surface area of the particle, $N_{\rm par} = {( V_{\rm par} N_{\rm A} \gamma_{\rm grain} )} / \Omega$ is the number of oxygen atoms in the particle, and $V_{\rm par} = {( 4 \pi / 3 )} {r_{\rm par}}^{3}$ is the volume of the particle.
The isotopic exchange efficiency of colliding molecules with particles, $\beta_{\rm par}$, is determined from laboratory experiments.
For amorphous forsterite grains reacting with water vapor, \citet{2018ApJ...865...98Y} measured the $\beta_{\rm par}$ value at $T \sim 800$--$900~\si{K}$ and $P = 10^{-2}~\si{Pa}$.
They found that
\begin{equation}
\beta_{\rm par} = 7.4 \times 10^{-6}.
\end{equation}

In this study, we neglect the temperature and pressure dependences of $\beta_{\rm par}$ for amorphous forsterite grains, although it would slightly depend on both $T$ and $P$ in reality \citep[e.g.,][]{YAMAMOTO202493}.
We also assume that $\beta_{\rm par}$ for crystalline forsterite grains is the same as that for amorphous grains, although in reality, it might be significantly lower than what we assumed for simplicity.

\subsection{Diffusion Within Constituent Particles}

For a spherical particle with a radius of $r_{\rm par}$, the timescale of the diffusive exchange reaction, $t_{\rm par, diff}$, is given by \citep[e.g.,][]{crank1979mathematics}
\begin{equation}
t_{\rm par, diff} = \frac{{r_{\rm par}}^{2}}{\pi^{2} D_{\rm par}},
\label{eq:t_par_diff}
\end{equation}
where $D_{\rm par}$ is the diffusive isotope exchange coefficient.
The temperature dependence of $D_{\rm par}$ follows the Arrhenius law: 
\begin{equation}
D_{\rm par} = D_{\rm par, ref} \exp{\left[ - \frac{E_{\rm a}}{R_{\rm gas}} {\left( \frac{1}{T} - \frac{1}{T_{\rm ref}} \right)} \right]},
\label{eq:Arrhenius}
\end{equation}
where $R_{\rm gas} = 8.31~\si{J.mol^{-1}.K^{-1}}$ is the gas constant, $E_{\rm a}$ is the activation energy, and $D_{\rm par, ref}$ is the diffusive isotope exchange coefficient at the reference temperature, $T_{\rm ref}$.

For amorphous forsterite grains in the ambient water vapor, \citet{2018ApJ...865...98Y} obtained the following values:
\begin{eqnarray}
D_{\rm par, ref} & = & 1.5 \times 10^{-19}~\si{m^{2}.s^{-1}}, \nonumber \\
T_{\rm ref}      & = & 1200~\si{K}, \nonumber \\
E_{\rm a}        & = & 161.5~\si{kJ.mol^{-1}}.
\end{eqnarray}
For crystalline forsterite grains, the temperature dependence of $D_{\rm par}$ is given by Equation (\ref{eq:Arrhenius}) with the following constants\footnote{
The mechanism for the diffusive isotope exchange within grains depends on whether the grains are amorphous or crystalline.
It could also depend on the molecular species of the ambient vapor \citep[see also][]{2010ChGeo.279....1M, 2018AmMin.103..412K, 2019AmMin.104..385K, 2019GeCoA.259..129Y}.} \citep{1980E&PSL..47..391J}:
\begin{eqnarray}
D_{\rm par, ref} & = & 3.6 \times 10^{-18}~\si{m^{2}.s^{-1}}, \nonumber \\
T_{\rm ref}      & = & 1823~\si{K}, \nonumber \\
E_{\rm a}        & = & 320~\si{kJ.mol^{-1}}.
\end{eqnarray}

\section{Results \& Discussion}
\label{sec:results}

We calculate the timescale for oxygen isotope exchange between dust aggregates and ambient gas, $t_{\rm ex}$, and investigate which process is the rate-limiting step.
For simplicity, we assume that dust aggregates are made of monodisperse spherical particles with radius of $r_{\rm par} = 0.1~\si{\micro m}$ \citep[e.g.,][]{2023ApJ...944L..43T}.
The volume filling factor of aggregates is set to $\phi_{\rm agg} = 0.3$.\footnote{
The choice of $\phi_{\rm agg} = 0.3$ is motivated from numerical simulations \citep[e.g.,][]{2010A&A...513A..57Z, 2023ApJ...951L..16A} and laboratory experiments \citep[e.g.,][]{2009ApJ...696.2036W, 2013Icar..225...75K} on collisional growth of dust aggregates.}
In this section, we consider oxygen isotope exchange with water vapor with partial pressure of $P = 10^{-4}~\si{Pa}$.

\begin{figure}
\centering
\includegraphics[width = \columnwidth]{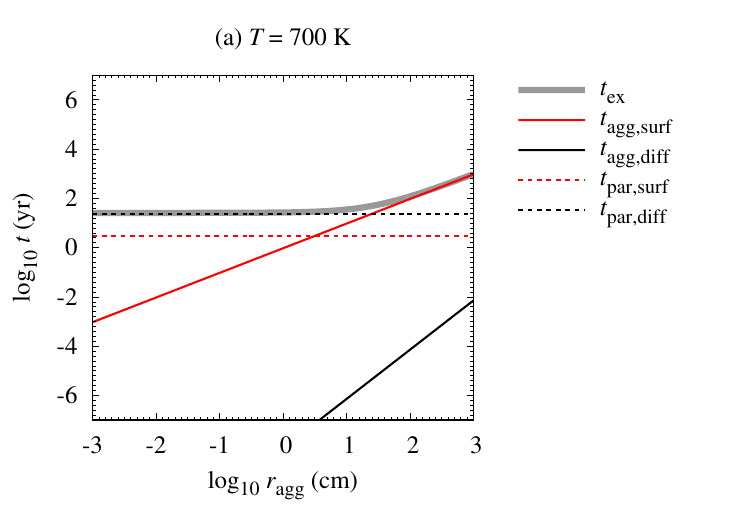}
\includegraphics[width = \columnwidth]{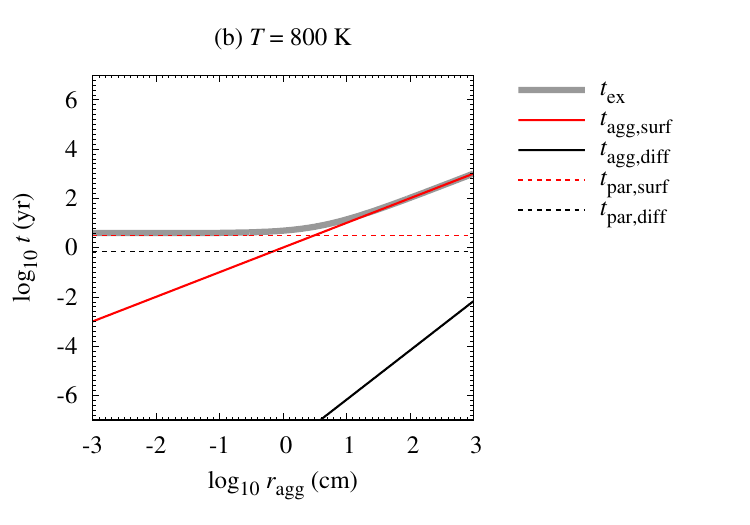}
\caption{
Timescale for oxygen isotope exchange between dust aggregates and ambient nebular gas, $t_{\rm ex}$, as a function of aggregate radius, $r_{\rm agg}$.
(a) For $T = 700~\si{K}$.
(b) For $T = 800~\si{K}$.
Here, we assume that aggregates are made of amorphous forsterite grains.
The timescales for four elementary processes ($t_{\rm agg, surf}$, $t_{\rm agg, diff}$, $t_{\rm par, surf}$, and $t_{\rm par, diff}$) are also plotted in the figure (see legends).
}
\label{fig_size}
\end{figure}

Figure \ref{fig_size} shows $t_{\rm ex}$ as a function of $r_{\rm agg}$.
Here, we assume that aggregates are made of amorphous forsterite grains.
It is evident that $t_{\rm agg, surf}$ is proportional to $r_{\rm agg}$ and $t_{\rm agg, diff}$ is proportional to ${r_{\rm agg}}^{2}$ (see Equations (\ref{eq:t_agg_surf}) and (\ref{eq:t_agg_diff}), respectively).
In contrast, $t_{\rm par, surf}$ and $t_{\rm par, diff}$ are independent of $r_{\rm agg}$.
We also reveal that $t_{\rm agg, diff}$ is orders of magnitude shorter than the other three timescales.
In other words, diffusion of molecules within aggregates is not the rate-limiting step.

Both $t_{\rm par, diff}$ and $t_{\rm par, surf}$ depend on $T$, and $t_{\rm par, diff} > t_{\rm par, surf}$ when $T = 700~\si{K}$ (see Figure \ref{fig_size}(a)).
In this case, either $t_{\rm par, diff}$ or $t_{\rm agg, surf}$ is the rate-limiting step: $t_{\rm ex} \simeq t_{\rm par, diff}$ when $r_{\rm agg} < 24~\si{cm}$, whereas $t_{\rm ex} \simeq t_{\rm agg, surf}$ when $r_{\rm agg} > 24~\si{cm}$.
In contrast, for $T = 800~\si{K}$ (see Figure \ref{fig_size}(b)), $t_{\rm par, surf} > t_{\rm par, diff}$, thus either $t_{\rm par, surf}$ or $t_{\rm agg, surf}$ is the rate-limiting step: $t_{\rm ex} \simeq t_{\rm par, surf}$ when $r_{\rm agg} < 3.2~\si{cm}$, whereas $t_{\rm ex} \simeq t_{\rm agg, surf}$ when $r_{\rm agg} > 3.2~\si{cm}$.

\begin{figure*}
\centering
\includegraphics[width = \columnwidth]{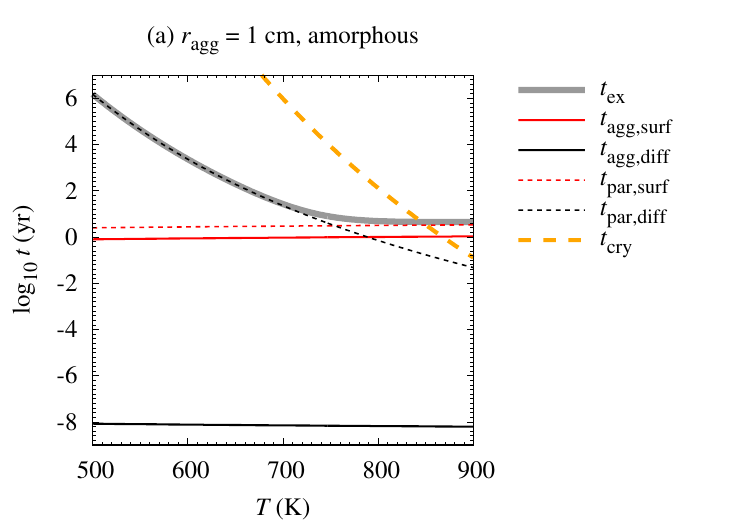}
\includegraphics[width = \columnwidth]{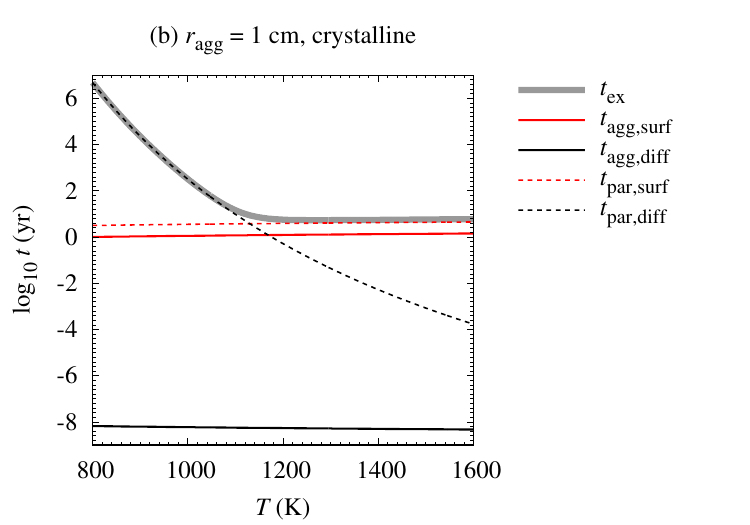}
\includegraphics[width = \columnwidth]{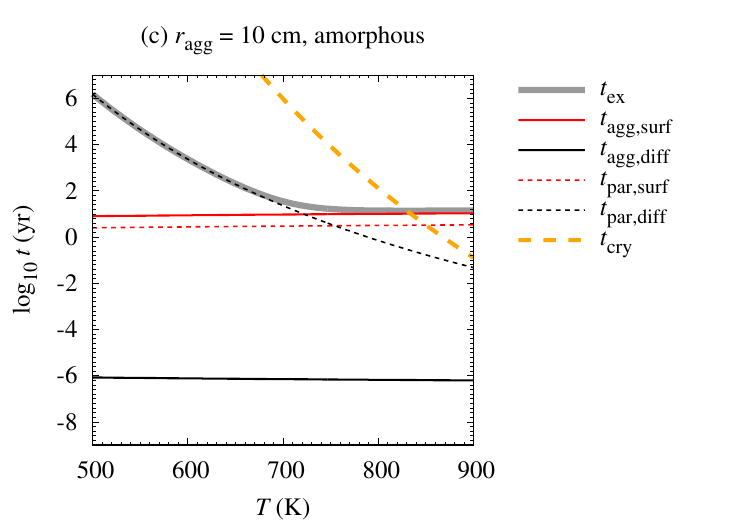}
\includegraphics[width = \columnwidth]{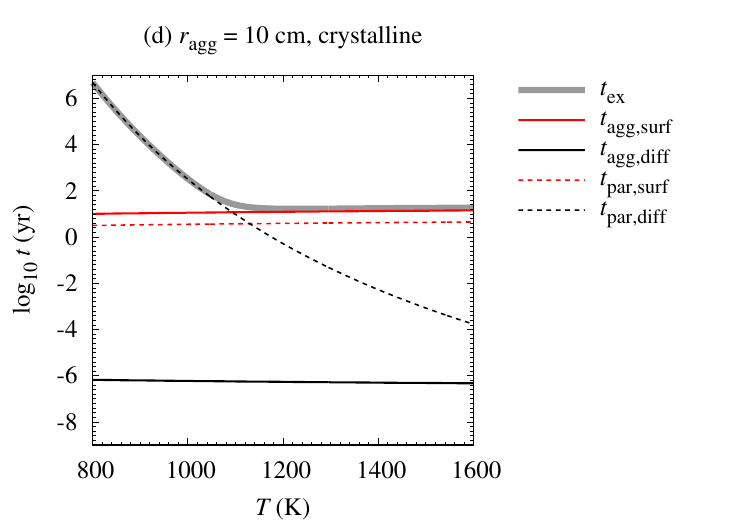}
\caption{
Timescale for oxygen isotope exchange between dust aggregates and ambient nebular gas, $t_{\rm ex}$, as a function of temperature, $T$.
(a) For $r_{\rm agg} = 1~\si{cm}$ and aggregates are made of amorphous forsterite particles.
(b) For $r_{\rm agg} = 1~\si{cm}$ and aggregates are made of crystalline forsterite particles.
(c) For $r_{\rm agg} = 10~\si{cm}$ and aggregates are made of amorphous forsterite particles.
(d) For $r_{\rm agg} = 10~\si{cm}$ and aggregates are made of crystalline forsterite particles.
The timescales for four elementary processes ($t_{\rm agg, surf}$, $t_{\rm agg, diff}$, $t_{\rm par, surf}$, and $t_{\rm par, diff}$) and for crystallization ($t_{\rm cry}$) are also plotted in the figure (see legends).
}
\label{fig_temp}
\end{figure*}

We also show the temperature dependence of $t_{\rm ex}$ in Figure \ref{fig_temp}.
Here, we calculate $t_{\rm ex}$ for both amorphous and crystalline cases, and $r_{\rm agg}$ is set to 1 cm or 10 cm.
Figure \ref{fig_temp}(a) shows the results for $r_{\rm agg} = 1~\si{cm}$ and aggregates are made of amorphous forsterite particles.
We find that $t_{\rm ex} \simeq t_{\rm par, diff}$ when $T < 750~\si{K}$, whereas $t_{\rm ex} \simeq t_{\rm par, surf}$ when $T > 750~\si{K}$.
In this case, $\max{( t_{\rm agg, surf}, t_{\rm agg, diff} )}$ is smaller than $t_{\rm ex, par}$ at an arbitrary temperature, and the impact of aggregate structure on the isotopic exchange timescale is negligible.

Amorphous silicate grains are crystallized under high temperature conditions by thermal annealing \citep[e.g.,][]{2000A&A...364..282F}.
The timescale of crystallization, $t_{\rm cry}$, is given by
\begin{equation}
t_{\rm cry} = \frac{1}{\nu_{0}} \exp{\left( \frac{E_{\rm cry}}{R_{\rm gas} T} \right)},
\end{equation}
where $\nu_{0}$ is the pre-exponential factor of the crystallization rate and $E_{\rm cry}$ is the activation energy for crystallization.
For amorphous forsterite grains with the ambient water vapor pressure of $P = 10^{-4}~\si{Pa}$, \citet{2018ESC.....2..778Y} found that $\ln{( \nu_{0} / {1~\si{Hz}} )} = 40.2$ and $E_{\rm cry} = 414.4~\si{kJ.mol^{-1}}$.
As shown in Figure \ref{fig_temp}(a), $t_{\rm cry}$ drastically decreases with increasing $T$, and $t_{\rm cry} = 1~\si{Myr}$ at $T = 700~\si{K}$.
Thus, a significant fraction of silicate dust particles in a high temperature region of $T > 700~\si{K}$ would be crystallized within a lifetime of protoplanetary disk. 
The dynamics of dust aggregates also affect the crystallization temperature \citep[see][]{2023ApJ...957...47I}.

Figure \ref{fig_temp}(b) shows the temperature dependence of $t_{\rm ex}$ for the case of $r_{\rm agg} = 1~\si{cm}$ and aggregates are made of crystalline forsterite particles.
As is the case for Figure \ref{fig_temp}(a), $\max{( t_{\rm agg, surf}, t_{\rm agg, diff} )}$ is always smaller than $t_{\rm ex, par}$, and $t_{\rm ex}$ is approximately given by $t_{\rm ex} \approx t_{\rm ex, par}$.

To understand the condition where the impact of aggregate structure on the isotopic exchange timescale becomes important, we compare $t_{\rm agg, surf}$ with $t_{\rm par, surf}$.
The ratio of the two timescales, $t_{\rm agg, surf} / t_{\rm par, surf}$, is given by
\begin{eqnarray}
\frac{t_{\rm agg, surf}}{t_{\rm par, surf}} & = & \frac{\beta_{\rm par} \phi_{\rm agg} r_{\rm agg}}{{( 1 - \phi_{\rm agg} )} r_{\rm par}} \nonumber \\
                                            & = & 0.32 {\left( \frac{\beta_{\rm par}}{7.4 \times 10^{-6}} \right)} {\left( \frac{r_{\rm agg}}{1~\si{cm}} \right)}.
\label{eq:t_t}
\end{eqnarray}
Therefore, $t_{\rm agg, surf}$ becomes larger than $t_{\rm par, surf}$ when $r_{\rm agg} > 3~\si{cm}$.
We validate this prediction in Figures \ref{fig_temp}(c) and \ref{fig_temp}(d).
For $r_{\rm agg} = 10~\si{cm}$, $t_{\rm agg, surf}$ is indeed larger than $t_{\rm par, surf}$, and $t_{\rm agg, surf}$ becomes the rate-limiting process at a high temperature ($T > 720~\si{K}$ for the amorphous case and $T > 1090~\si{K}$ for the crystalline case).
We note that $t_{\rm agg, surf} / t_{\rm par, surf}$ is independent of $\Omega$ and $\gamma_{\rm grain}$.

Multi-wavelength observations of protoplanetary disks at (sub)millimeter wavelengths allow us to constrain the size of dust aggregates and its radial distribution \citep[e.g.,][]{2019ApJ...883...71C, 2021ApJ...913..117U}.
Several studies \citep[e.g.,][]{2014A&A...564A..93M, 2015ApJ...813...41P, 2016A&A...588A..53T} have reported that large aggregates with $r_{\rm agg} \gtrsim 1~\rm{cm}$ are ubiquitous in the inner parts of disks where $r \lesssim 10~\si{au}$, whereas the maximum radius of aggregates is typically a few millimeters in the outer parts of $r \gtrsim 100~\si{au}$.
Comets in the solar system also consist of millimeter- to decimeter-sized dust aggregates referred to as ``pebbles'' \citep[e.g.,][]{2020MNRAS.497.1166A, 2022Univ....8..381B}.
These pieces of evidence support the presence of large dust aggregates in the ancient solar nebula.

In the protosolar molecular cloud, $^{16}$O-poor H$_{2}$O reservoir would be formed through the self-shielding of CO gas, and silicates and volatiles have different oxygen isotopic compositions at the earliest stage of the solar nebula \citep[e.g.,][]{2020SciA....6.2724K}.
Centimeter-sized icy dust aggregates drift toward the Sun due to gas drag \citep[e.g.,][]{1976PThPh..56.1756A, 1977MNRAS.180...57W}, although the direction of radial migration depends on the sign of the pressure gradient \citep[e.g.,][]{2021ApJ...920...27A}.
Furthermore, the radial migration of icy dust aggregates, followed by their evaporation at the snow line, could cause the spatiotemporal variation of oxygen isotopic composition in the gas component of the inner solar nebula \citep[e.g.,][]{2004ApJ...614..490C, 2005ApJ...622.1333K}.

In this study, we derived the fundamental equations for the isotope exchange process between aggregates and ambient vapor (Equations (\ref{eq:t_agg_surf}) and (\ref{eq:t_agg_diff})).
Using these equations, we plan to investigate the spatiotemporal evolution of oxygen isotopic composition and silicate crystallinity, in conjunction with disk formation and evolution, in future studies.

\section{Conclusion}

In this study, we theoretically estimated the oxygen isotope exchange timescales between silicate dust aggregates and ambient water vapor.
The isotope exchange process between aggregates and ambient vapor is divided into four processes (Section \ref{sec:model}).
We evaluated the timescales of these processes and assess which one becomes the rate-determining step.
Our analytical calculations revealed that the isotope exchange timescale is approximately the same as that of the constituent particles when the aggregate radius is smaller than the critical value given by Equation (\ref{eq:t_t}) (Section \ref{sec:results}).
We plan to perform numerical simulations of oxygen isotopic evolution of the solar nebula using the theoretical model derived in this study.


\section*{Acknowledgments}

This work was supported by JSPS KAKENHI Grant Numbers JP24K17118 and JP21K13986.


%




\bibliography{sample631}{}
\bibliographystyle{aasjournal}



\end{document}